# Lattice energy-momentum tensor
# with Symanzik improved actions.

Alessandra Buonanno[1]

*Dipartimento di Fisica dell'Università, Piazza Torricelli 2, I-56126 Pisa, Italy*

Giancarlo Cella[2]

*Dipartimento di Fisica dell'Università, Piazza Torricelli 2, I-56126 Pisa, Italy, and Istituto Nazionale di Fisica Nucleare, Via Livornese 582/a I-56010 S. Piero a Grado (Pisa), Italy*

Giuseppe Curci[3]

*Dipartimento di Fisica dell'Università, Piazza Torricelli 2, I-56126 Pisa, Italy, and Istituto Nazionale di Fisica Nucleare, Via Livornese 582/a I-56010 S. Piero a Grado (Pisa), Italy*

(September 6, 1994)

## Abstract

We define the energy-momentum tensor on lattice for the $\lambda \phi^4$ and for the nonlinear $\sigma$-model Symanzik tree-improved actions, using Ward identities or an explicit matching procedure.

The resulting operators give the correct one loop scale anomaly, and in the case of the sigma model they can have applications in Monte Carlo simulations.

PACS numbers: 11.10.Gh, 11.30.Cp

Typeset using REVTeX

# I. INTRODUCTION.

The lattice regularization is the preferred framework for the non perturbative study of a field theory by means of numerical simulations. It gives the bonus of explicit preservation of gauge invariance, which is believed to be the key to understand very important phenomena, as for example confinement. The lattice however breaks other symmetries, as the translation (Poincaré) invariance, and it is important to verify the correct restoration of these properties in the continuum limit.

A classical symmetry is expressed at quantum level by a series of Ward identities involving the associated current operator. These identities can't be satisfied in presence of a non invariant regulator $\Lambda$. If the invariance must be preserved, we can expect that, after a suitable $O(\hbar)$ operator redefinition, these relations will be broken only by "irrelevant" (that is to say, which vanish removing the regularization) terms.

In general this program cannot be accomplished, as a classical invariance may not be compatible with the quantization. A well known example is the dilatation invariance, that independently of the regularization chosen cannot be restored in the renormalized theory. However, if an invariant regulator can be found, we must be able to restore the Ward identities in another scheme, whatever it would be.

This is the case of the translation invariance, for which an invariant scheme exists (for example the dimensional regularization). The relative Ward identities are relations between insertions of the energy-momentum tensor (EMT) that can be written as

$$\partial_\mu^x \Gamma_{T_{\mu\nu}}^{(n)}(x; x_1, \cdots, x_n) = -\sum_{i=1}^{n} \partial_\nu^x \delta^{(4)}(x - x_i) \Gamma^{(n)}(x_1, \cdots, x_n). \tag{1}$$

On the lattice, we can perturbatively impose the validity of these equations up to terms $O(a)$ (or $O(a^2)$ in absence of fermionic fields): explicit one loop calculations have been done for the $\lambda\phi^4$ model [4], for the QED [5], and for the QCD [6,7].

In a concrete lattice simulation the cutoff $\Lambda = 1/a$ is obviously finite. This gives rise to systematic errors in the estimation of observable quantities that can be reduced using very large lattices. An alternative possibility, proposed by Symanzik [8] is to redefine the lattice action: we can exploit the ambiguity inherent to the lattice transcription and perturbatively eliminate the cutoff effects to a given power of $a$.

We have explicitly calculated the finite corrections necessary for the one loop definition of the EMT with a Symanzik tree improved action. First, as an academic exercise, we have analysed the simple $\lambda\phi^4$ model [9]. Next we have considered the nonlinear $O(N)$ bidimensional $\sigma$ model [10]. This is of much interest as it shares some important properties with lattice gauge theories [11], first of all the asymptotic freedom which is an argument to justify a perturbative approach.

Since the EMT is related with the dilatation current, which as we said is broken at quantum level, we have also verified that the so defined operator gives the correct one loop anomaly.

## II. SYMANZIK IMPROVEMENT.

We use the $\lambda\phi^4$ model to explain in detail the procedure. Starting from the classical continuum action $S$ we can write a lattice regularized version of it $S_{latt}$ in infinite ways, requiring only that

$$\lim_{a \to 0} S_{latt}[\phi] = S[\phi]. \qquad (2)$$

The most simple option is the substitution of the integral with a sum over lattice sites $n$, and of the derivatives with the finite differences $\Delta^+_\mu \phi(n) = [\phi(n + a\hat{\mu}) - \phi(n)]/a$,

$$S^{(0,0)}_{latt}[\phi] = a^4 \sum_n \left[ \frac{1}{2} \sum_\mu \Delta^+_\mu \phi(n) \Delta^+_\mu \phi(n) + \frac{1}{2} m^2 \phi^2(n) + \frac{1}{4!} g \phi^4(n) \right] = S[\phi] + O(a^2). \qquad (3)$$

If we limit ourselves to a tree level (classical) improvement we can easily do better. Differences proportional to powers of $a$ between the two actions arise from the derivative term, and can be compensated adding to the lattice lagrangian suitable irrelevant operators, or equivalently using an improved finite difference formally defined by

$$\partial_\mu = \frac{1}{a} \log \left( 1 + a \Delta^+_\mu \right) \qquad (4)$$

with the second term expanded at the desired order. The classical action with the $O(a^2)$ artifacts removed is given by

$$S^{(1,0)}_{latt}[\phi] = S^{(0,0)}_{latt}[\phi] + a^4 \sum_{n,\mu} \frac{1}{4!} a^2 \Delta^+_\mu \Delta^+_\mu \phi(n) \Delta^+_\mu \Delta^+_\mu \phi(n) = S[\phi] + O(a^4), \qquad (5)$$

as can be easily verified expanding the finite differences.

In the quantized theory it is necessary to parametrize the divergences of the continuum theory, for instance using dimensional regularization, and to define an improvement criterion. We require that the lattice one particle irreducible functions $\Gamma$ generated by $S^{(n,l)}_{latt}$ coincide at small external momenta with the continuum ones, apart from $o(\hbar^l) + o(a^{2n})$ corrections. Symanzik has shown that it is possible to write

$$S^{(n,l)}_{latt}[\phi] = S^{(n,0)}_{latt}[\phi] + \sum_{k=0}^{n} \sum_{p=1}^{l} \hbar^p \sum_j^{N_k} f_{jp}(g) \left( a^D \sum_n a^{2k} \mathcal{O}^{(2k+D)}_j(n) \right), \qquad (6)$$

where the $\mathcal{O}^{(k)}$ are k-dimensional lattice operators, and the coefficients $f_{jp}$ are calculable imposing the improvement criterion with a $p$ loops calculation.

### A. Energy momentum tensor.

From now on we work in the tree improved theory defined by $S^{(1,0)}_{latt}$, which is equivalent to an effective continuum (dimensionally regularized) action that can be written as

$$S_{eff} = \int d^D x \left[ \frac{1}{2} Z_\phi \partial_\mu \phi \partial_\mu \phi + \frac{1}{2} m^2 Z_m \phi^2 + \frac{1}{4!} g Z_g \phi^4 \right]. \qquad (7)$$

The $Z$ constant can be determined imposing the equivalence of superficially divergent $\Gamma$ functions. For the two points 1PI function we have on lattice (see app. A for the notations)

$$\Gamma^{(2)}_{latt}(0,0) = m^2 + \frac{g}{2} \int \frac{d^4 l}{(2\pi)^4} D_{latt}(l) = \frac{g}{2} \left[ \frac{Z_0}{a^2} + \frac{m^2}{(4\pi)^2} \left( \log a^2 m^2 - F_{0000} - 1 + \gamma_E - T_0 \right) \right] \tag{8}$$

and in the continuum

$$\Gamma^{(2)}_{cont}(0,0) = m^2 + \frac{g\mu^{4-D}}{2} \int \frac{d^D l}{(2\pi)^D} D_{cont}(l) = m^2 + \frac{g}{2} \left[ \frac{m^2}{(4\pi)^2} \left( -\frac{1}{\epsilon} - 1 + \gamma_E + \log \frac{m^2}{4\pi\mu^2} \right) \right]. \tag{9}$$

Now we must subtract the pole part in (9) (we use minimal subtraction in the continuum), and dispose of the quadratically divergent piece in (8) by imposing the vanishing of self energy on lattice for $p = m = 0$ with a suitable counterterm

$$\Delta S^{(1,0)}_{latt} = -a^4 \sum_n g \frac{Z_0}{4a^2} \phi^2(n). \tag{10}$$

Comparing the two results we obtain (we set the mass scale as $\mu = 1/a$ for simplicity)

$$Z_m = 1 + \frac{g}{2(4\pi)^2} \left[ \log 4\pi - T_0 - F_{0000} \right]. \tag{11}$$

The procedure for the determination of $Z_g$ is analogous. For the four points functions we obtain on lattice

$$\Gamma^{(4)}_{latt}(0,0,0,0) = g - \frac{3}{2} g^2 \int \frac{d^4 l}{(2\pi)^4} D^2_{latt}(l) = g - \frac{3}{2} \frac{g^2}{(4\pi)^2} \left[ -\gamma_E + F_{0000} + T_0 - \log a^2 m^2 \right], \tag{12}$$

and in the continuum

$$\Gamma^{(4)}_{cont}(0,0,0,0) = g - \frac{3}{2} g^2 \int \frac{d^D l}{(2\pi)^D} D^2_{cont}(l) = g - \frac{3}{2} \frac{g^2}{(4\pi)^2} \left[ \frac{1}{\epsilon} - \gamma_E - \log \frac{m^2}{4\pi\mu^2} \right] \tag{13}$$

so that

$$Z_g = 1 + \frac{3}{2} \frac{g^2}{(4\pi)^2} \left[ \log 4\pi - T_0 - F_{0000} \right]. \tag{14}$$

It is now possible to define the EMT in a very simple way. In fact it can be shown [12,13] that, given the effective lagrangian $L_{eff}$ defined by (7), it is sufficient to write a lattice operator which reduces in the leading order in $a$ to

$$T^{eff}_{\mu\nu} = Z_\phi N[\partial_\mu \phi \partial_\mu \phi] - \delta_{\mu\nu} N[L_{eff}], \tag{15}$$

where $N[O]$ is an operator corrected accordingly to the prescription of the effective lagrangian. We note that as we consider only the integrated $T_{\mu\nu}$, there are no problems of ambiguity in the definition of the classical operator, so that we can neglect the so-called "improved Coleman term" [17,16].

We can also write at one loop order

$$T_{\mu\nu} = Z_1 \bar{\Delta}_\mu \phi \bar{\Delta}_\nu \phi - \frac{a^2}{6}\left(\bar{\Delta}_\mu \phi \bar{\Delta}_\nu^3 \phi + \bar{\Delta}_\mu^3 \phi \bar{\Delta}_\nu \phi\right) +$$
$$- \delta_{\mu\nu}\left[\frac{Z_2}{2}\sum_\rho \bar{\Delta}_\rho \phi \bar{\Delta}_\rho \phi - \frac{a^2}{6}\sum_\rho \bar{\Delta}_\rho \phi \bar{\Delta}_\rho^3 \phi + \frac{Z_3}{2}\bar{\Delta}_\mu \phi \bar{\Delta}_\mu \phi + \frac{Z_4}{2}m^2\phi^2 + \frac{Z_5}{4!}g\phi^4\right], \quad (16)$$

where we have used the symmetric finite difference $\bar{\Delta}_\mu \phi(n) = (\phi(n+a\hat{\mu}) - \phi(n-a\hat{\mu}))/2a$, and we have imposed the Ward identities to determine the $Z$ finite corrections. We have explicitly checked that the two methods agree each other.

The expression (16) reduces to the classical improved quantity if $Z_1 = Z_2 = Z_4 = Z_5 = 1, Z_3 = 0$, and as we are considering only tree improvement we can neglect $O(a^2 g)$ corrections. The resulting Feynman rules give two and four lines vertices, $T^{(2)}_{\mu\nu}$ and $T^{(4)}_{\mu\nu}$. The relevant identities defined by (1) involve the insertion of these vertices in the two and four points irreducible functions (see figure (1)). We start from the two points identity, written in momentum space for the integrated $T_{\mu\nu}$:

$$\Gamma^{(2)}_{T_{\mu\nu}}(0; p, -p) = \left(\frac{p_\mu}{2}\frac{\partial}{\partial p_\nu} + \frac{p_\nu}{2}\frac{\partial}{\partial p_\mu} - \delta_{\mu\nu}\right)\Gamma^{(2)}(p). \quad (17)$$

We have already calculated the one loop correction to $\Gamma^{(2)}$ (see eq. (8)), while for the insertion we obtain

$$\Gamma^{(2)}_{T_{\mu\nu}}(0; p, -p) = 2T^{(2)}_{\mu\nu}(p, -p) - \frac{g}{2}\int \frac{d^4 l}{(2\pi)^4} D_{latt}(l)\left[\delta_{\mu\nu} + 2T^{(2)}_{\mu\nu}(l, -l)D_{latt}(l)\right], \quad (18)$$

and substituing in the (17) we find the conditions

$$(Z_1 - 1) = (Z_2 - 1) = Z_3 = O(g^2) \quad (19)$$

$$Z_4 = 1 + \frac{g}{4}\int \frac{d^4 l}{(2\pi)^4} D^2_{latt}(l)\left[2 + \frac{\bar{l}^2}{m^2} + \frac{a^2}{3}\frac{\bar{l}^4}{m^2}\right] =$$
$$= 1 + \frac{g}{2(4\pi)^2}\left[T_0 - T_2 + 2\pi^2 Z_{0000} - 1 - \frac{1}{3}T_4\right] + \frac{g}{a^2 m^2}\left[\frac{T_1}{4} + \frac{T_3}{12}\right]. \quad (20)$$

The determination of the $Z_5$ correction can be done considering the four points identity

$$\Gamma^{(4)}_{T_{\mu\nu}}(0; 0, 0, 0, 0) = -\delta_{\mu\nu}\Gamma^{(4)}(0, 0, 0, 0). \quad (21)$$

We calculate the relevant insertion which, taking into account the relations (20), is of the form

$$\Gamma^{(4)}_{T_{\mu\nu}}(0; 0, 0, 0, 0) = -\delta_{\mu\nu} g Z_5 + 3g^2 \delta_{\mu\nu}\int \frac{d^4 l}{(2\pi)^4} D^2_{latt}(l) +$$
$$+3g^2 \int \frac{d^4 l}{(2\pi)^4} D^3_{latt}(l)\left[2\bar{l}_\mu \bar{l}_\nu + \frac{a^2}{3}\left(\bar{l}_\mu \bar{l}_\nu^3 + \bar{l}_\nu \bar{l}_\mu^3\right) - \delta_{\mu\nu}\left(\bar{l}^2 + \frac{a^2}{3}\bar{l}^4 + m^2\right)\right]. \quad (22)$$

Using this result and the eq. (12) we obtain

$$Z_5 = 1 + \frac{3}{2}g \int \frac{d^4l}{(2\pi)^4} D_{latt}^2(l) \left[ 1 - D_{latt}(l) \left( \tilde{l}^2 + 2m^2 + \frac{a^2}{3}l^4 \right) \right] =$$
$$= 1 + 3\frac{g}{2(4\pi)^2} \left[ T_0 - T_2 + 2\pi^2 Z_{0000} - 1 - \frac{1}{3}T_4 \right]. \quad (23)$$

### B. The dilatation current anomaly.

From the Ward identity for the Noëther current connected to scale invariance

$$J_\mu^{(D)} = x_\nu T_{\mu\nu} \quad (24)$$

one can show that the scaling equation is related to the EMT trace as follows

$$\left( a\frac{\partial}{\partial a} - m\frac{\partial}{\partial m} - N \right) \Gamma^{(N)}(p_1, \cdots, p_N) \simeq \Gamma^{(N)}_{T_{\mu\mu}}(0; p_1, \cdots, p_N) \quad (25)$$

Now, the mass derivative is equivalent to the insertion in $\Gamma$ of a renormalized mass operator, and analogously the field number $N$ can be generated inserting the renormalized operator $\phi\frac{\delta S}{\delta \phi}$. It follows that scale anomaly is equivalent to the insertion of the evanescent operator

$$a^4 \sum_n \left( T_{\mu\mu} + m^2 \phi^2 + \phi \frac{\delta S_{latt}}{\delta \phi} \right). \quad (26)$$

Using the defined EMT we obtain the same results one finds with the not improved action [4]. This is correct, as the anomaly is proportional to the $\beta$ functions of the theory that, as well known, does not depend to this order from the regularization scheme.

### III. RESULTS FOR THE IMPROVED NONLINEAR SIGMA MODEL.

The nonlinear $O(N)$ sigma model is defined by the path integral

$$Z = \int \mathcal{D}\vec{\phi}(x)\, \delta(\vec{\phi}^2(x) - 1) \exp\left\{ -\frac{1}{2t} \int d^2y\, \partial_\mu \vec{\phi}(y) \cdot \partial_\mu \vec{\phi}(y) \right\}, \quad (27)$$

where $\vec{\phi} = (\phi_1, \cdots, \phi_N)$ are $N$ scalar fields. In the continuum we use as usual the dimensional regularization scheme, with minimal subtraction at the scale $\mu = 1/a$, and in order to eliminate infrared divergences, which are an artefact of perturbative approximation, we add to the action a mass term which breaks the $O(N)$ symmetry. With the chosen regularization there are no contributions from the path integral measure, and we can write the action as

$$S = \int d^D x\, \mu^{D-2} \left[ \frac{1}{2t} \partial_\mu \vec{\pi} \cdot \partial_\mu \vec{\pi} + \frac{1}{2t} \partial_\mu \sigma \partial_\mu \sigma - \frac{m^2}{t} \sigma \right], \quad (28)$$

with the parametrization

$$\vec{\pi} = (\phi_1, \cdots, \phi_{N-1})$$
$$\sigma = \phi_N = \sqrt{1 - \vec{\pi} \cdot \vec{\pi}}. \tag{29}$$

This lagrangian generates an infinite number of vertices, but at a given perturbative order only a restricted set of them gives a contribution. At one loop we need the propagator of $\pi_i$ fields $D_{ij}(p)$ and the four points vertex $V_{ijkl}(p_1, p_2, p_3, p_4)$.

The $O(N)$ invariance gives strong constraints to the renormalized action. Solving explicitly the relative Ward identities [14] one can prove that this can be written in terms of two renormalization constant $Z, Z_t$,

$$S^{(R)} = \int d^D x\, \mu^{D-2} \left[ \frac{Z}{2Z_t t} (\partial_\mu \vec{\pi} \cdot \partial_\mu \vec{\pi} + \partial_\mu \sigma \partial_\mu \sigma) - \frac{m^2}{t} \sigma \right],$$
$$\sigma = \sqrt{Z^{-1} - \vec{\pi}^2} \tag{30}$$

On lattice the tree improved action can be written as

$$S^{(1,0)}_{latt} = S_0 + S_1 + S_{meas} \tag{31}$$

with

$$S_0 = a^2 \sum_n \left[ \frac{1}{2t} \Delta_\mu^+ \vec{\pi}(n) \cdot \Delta_\mu^+ \vec{\pi}(n) + \frac{1}{2t} \Delta_\mu^+ \sigma(n) \Delta_\mu^+ \sigma(n) - \frac{m^2}{t} \sigma(n) \right] \tag{32}$$

$$S_1 = a^2 \sum_n \left[ \frac{a^2}{24t} \Delta_\mu^+ \Delta_\mu^+ \vec{\pi}(n) \cdot \Delta_\mu^+ \Delta_\mu^+ \vec{\pi}(n) + \frac{a^2}{24t} \Delta_\mu^+ \Delta_\mu^+ \sigma(n) \Delta_\mu^+ \Delta_\mu^+ \sigma(n) \right] \tag{33}$$

$$S_{meas} = a^2 \sum_n \frac{1}{a^2} \log \sigma(n), \tag{34}$$

the last term coming from the exponentiated path integral measure. This action is equivalent to an effective continuum one, that must be of the form (30). To determine the finite renormalizations $Z, Z_t$ it is sufficient to compare the two point irreducible function on the continuum

$$\Gamma^{(2)}_{ij,cont}(p, -p) = \delta_{ij} \left[ \frac{p^2 + m^2}{t} - \frac{1}{4\pi} \left( p^2 + \frac{N-1}{2} m^2 \right) \left( \log \frac{a^2 m^2}{4\pi} + \gamma_E \right) \right] \tag{35}$$

and on the lattice

$$\Gamma^{(2)}_{ij,latt}(p, -p) = \frac{\delta_{ij}}{t} \left[ m^2 + \sum_\rho \left( \widehat{p}_\rho^2 + \frac{a^2}{12} \widehat{p}_\rho^4 \right) - \frac{t}{a^2} + t \frac{N+1}{2} m^2 \int \frac{d^2 l}{(2\pi)^2} D_{latt}(l) + \right.$$
$$\left. + \frac{t}{2} \int \frac{d^2 l}{(2\pi)^2} D_{latt}(l) \sum_\rho \left( \widehat{(p+l)}_\rho^2 + \widehat{(p-l)}_\rho^2 + \frac{a^2}{12} \widehat{(p+l)}_\rho^4 + \frac{a^2}{12} \widehat{(p-l)}_\rho^4 \right) \right]. \tag{36}$$

Using the lattice integrals properties and suppressing irrelevant terms we can write

$$\Gamma^{(2)}_{ij,latt}(p, -p) \simeq \frac{\delta_{ij}}{t} \left[ p^2 + m^2 + \right.$$
$$\left. + \int \frac{d^2 l}{(2\pi)^2} D_{latt}(l) \left( \frac{N-1}{2} m^2 + (1 + 4 \cos a l_1 - 2 \cos^2 a l_1) \frac{p^2}{3} \right) \right], \tag{37}$$

where the quadratic divergence in the integral has been canceled by the measure term. The final result is

$$Z_t = 1 + \frac{t}{4\pi}(N-2)\left(\gamma_E + E_0 + \log\frac{32}{4\pi}\right) + \frac{t}{4\pi}\left(\frac{8}{3} - \frac{4}{3}E_5 + \frac{2}{3}E_0 + \frac{2}{3}E_3\right) \tag{38}$$

$$Z = 1 + \frac{t}{4\pi}(N-1)\left(\gamma_E + E_0 + \log\frac{32}{4\pi}\right). \tag{39}$$

### A. Energy momentum tensor.

At first we note that we expect to obtain a $O(N)$ invariant result, except for contributions proportional to the symmetry breaking mass or to the motion equation

$$\frac{\delta S_{latt}}{\delta \vec{\pi}} = -\frac{1}{t}\left[\Delta_\mu^+ \Delta_\mu^- \vec{\pi} - \frac{\vec{\pi}}{\sigma}\Delta_\mu^+ \Delta_\mu^- \sigma - m^2 \frac{\vec{\pi}}{\sigma}\right]. \tag{40}$$

So at the order we are working the operator can be written as

$$T_{\mu\nu} = T_{\mu\nu}^{(0)} + T_{\mu\nu}^{(1)} + T_{\mu\nu}^{(m.e.)} + T_{\mu\nu}^{(l)} \tag{41}$$

with

$$T_{\mu\nu}^{(0)} = \frac{Z_1}{t}\bar{\Delta}_\mu\vec{\phi}\cdot\bar{\Delta}_\nu\vec{\phi} - \delta_{\mu\nu}\left[\frac{Z_2}{2t}\sum_\rho \bar{\Delta}_\rho\vec{\phi}\cdot\bar{\Delta}_\rho\vec{\phi} - \frac{Z_m}{t}m^2\sigma\right], \tag{42}$$

$$T_{\mu\nu}^{(1)} = -\frac{a^2}{6t}\left(\bar{\Delta}_\mu\vec{\phi}\cdot\bar{\Delta}_\nu^3\vec{\phi} + \bar{\Delta}_\mu^3\vec{\phi}\cdot\bar{\Delta}_\nu\vec{\phi}\right) + \delta_{\mu\nu}\frac{a^2}{6t}\sum_\rho \bar{\Delta}_\rho\vec{\phi}\cdot\bar{\Delta}_\rho^3\vec{\phi}, \tag{43}$$

$$T_{\mu\nu}^{(m.e.)} = Z_4\vec{\pi}\cdot\frac{\delta S_{latt}}{\delta\vec{\pi}}, \tag{44}$$

$$T_{\mu\nu}^{(l)} = -\delta_{\mu\nu}\frac{Z_3}{2t}\bar{\Delta}_\mu\vec{\phi}\cdot\bar{\Delta}_\mu\vec{\phi}. \tag{45}$$

When $Z_1 = Z_2 = 1$ the term $T_{\mu\nu}^{(0)}$ reduces to the naive lattice transcription of the classical operator. $T_{\mu\nu}^{(1)}$ cancels the $O(a^2)$ lattice artifacts at tree level, and we have not considered $O(t)$ corrections for it as they would be irrelevant to the order we are working. $T_{\mu\nu}^{(m.e.)}$ is a possible contribution proportional to the motion equation, and as we said it doesn't need to be $O(N)$ invariant. It is not present at tree level, so $Z_4 = O(t)$. $T_{\mu\nu}^{(l)}$ is an operator with only the hypercubic lattice symmetry that can be required to ripristinate the $O(a^2)$ rotational invariance.

As in the $\lambda\Phi^4$ model it is possible to calculate the $Z$ corrections by using two equivalent methods. We start evaluating the insertion with external momenta $p = k$ and $q = 0$ (see the graphs I0,I1,I2 in fig. (1)). In this regime the Ward identity reduces to

$$k_\nu \Gamma^{(2)}_{ij,T_{\mu\nu}}(k;k,0) = -k_\mu \Gamma^{(2)}_{ij}(0,0). \tag{46}$$

The contribution of the various insertions can be written as

$$I_0(k,0) = \delta_{ij}\delta_{\mu\nu}\frac{1}{t}\left[Z_4\left(\hat{k}^2 + 2m^2\right) - Z_m m^2\right] \tag{47}$$

$$I_1(k,0) = -\frac{\delta_{ij}}{2}\int\frac{d^2l}{(2\pi)^2}\left[2D_{latt}(l) + (N-1)\frac{D_{latt}(l)D_{latt}(l+k)}{D_{latt}(k)}\right] \times [A_{\mu\nu}(k+l,l)+$$
$$- \delta_{\mu\nu}\left(\frac{1}{2}A_{\rho\rho}(k+l,k) + m^2\right)] \tag{48}$$

$$I_2(k,0) = \delta_{ij}\int\frac{d^2l}{(2\pi)^2}D_{latt}(l)\left[A_{\mu\nu}(k+l,l) - \delta_{\mu\nu}\left(\frac{1}{2}A_{\rho\rho}(k+l,k) + \frac{N+1}{2}m^2\right)\right] \tag{49}$$

$$A_{\mu\nu}(p,q) = \overline{p}_\mu \overline{q}_\nu + \frac{a^2}{6}\left(\overline{p}_\mu^3 \overline{q}_\nu + \overline{p}_\mu \overline{q}_\nu^3\right) + [\mu \leftrightarrow \nu] \tag{50}$$

Summing the three contributions and using (46) we obtain

$$Z_4 = \frac{t}{8\pi}(N-1) \tag{51}$$

$$Z_m = 1 - \frac{t}{8\pi}(N-1). \tag{52}$$

The other constants can be calculated from the insertions of the integrated operator. With the same notation we have

$$I_0(p,-p) = \frac{\delta_{ij}}{t}\left[2Z_1\overline{p}_\mu\overline{p}_\nu + \frac{a^2}{3}\left(\overline{p}_\mu\overline{p}_\nu^3 + \overline{p}_\mu^3\overline{p}_\nu\right) - \delta_{\mu\nu}\left(Z_2\overline{p}^2 + \frac{a^2}{3}\overline{p}^4 + Z_3\overline{p}_\mu^2 + Z_m m^2 +\right.\right.$$
$$+ 2Z_4(\hat{p}^2 + m^2))] \tag{53}$$

$$I_1(p,-p) = \delta_{ij}\delta_{\mu\nu}\frac{m^2}{2}\int\frac{d^2l}{(2\pi)^2}D_{latt}^2(l)\left[2D_{latt}(p+l) + (N-1)m^2\right] \tag{54}$$

$$I_2(p,-p) = \delta_{ij}\int\frac{d^2l}{(2\pi)^2}D_{latt}(l)\left[2A_{\mu\nu}(p+l,p+l) - \frac{\delta_{\mu\nu}}{2}\left(A_{\rho\rho}(l+p,l+p) + (N+1)m^2\right)\right] \tag{55}$$

Taking all together, and using the Ward identity (17) we obtain

$$Z_1 = 1 + \frac{t}{12\pi}(4E_5 + E_0 - 3E_1 - 2E_3 - 3E_2 - 20) \tag{56}$$

$$Z_2 = 1 + \frac{t}{12\pi}(4E_5 - 2E_0 - 2E_3 + 8E_4 - 3(N-1) - 5) \tag{57}$$

$$Z_3 = \frac{t}{6\pi}(3E_0 - 3E_1 - 3E_2 - 8E_4 - 12). \tag{58}$$

This result can be checked by an explicit determination of a lattice transcription of the effective EMT

$$T_{\mu\nu}^{eff} = \frac{1}{t}\frac{Z}{Z_t}N[\partial_\mu\vec{\phi}\cdot\partial_\nu\vec{\phi}] - \delta_{\mu\nu}N[L_{eff}], \tag{59}$$

where $L_{eff}$ is defined by (30). It is sufficient to match lattice and continuum insertion of operators in the two points function.

For reference we list the results for the matching coefficients

$$b_1 = 1 + \frac{t}{4\pi}\left[-4 - E_2 - E_1 - \log\frac{32}{4\pi} + \gamma_E\right], \tag{60}$$

$$b_2 = \frac{t}{8\pi}\left[(N-2)(-\log\frac{32}{4\pi} - \gamma_E + E_6 + 1) - \frac{E_7}{24}\right], \tag{61}$$

$$b_3 = \frac{t}{12\pi}\left[12 - 3E_0 + 3E_1 + 3E_2 + 8E_4\right], \tag{62}$$

$$b_4 = b_5 = \frac{t}{8\pi}(N-1)\left[-1 - E_6 + \log\frac{32}{4\pi} + \gamma_E\right], \tag{63}$$

$$c_1 = 1 + \frac{t}{4\pi}\left[(N-2)E_6 - (N-1)\left(\log\frac{32}{4\pi} + \gamma_E\right) - \frac{1}{24}E_7 - E_0 + \frac{8}{3}E_4\right] \tag{64}$$

$$c_2 = c_3 = \frac{t}{4\pi}(N-1)\left[\log\frac{32}{4\pi} + \gamma_E - E_6\right], \tag{65}$$

$$\tag{66}$$

defined by

$$N[\partial_\mu\vec{\phi}\cdot\partial_\nu\vec{\phi}] = b_1\bar{\Delta}_\mu\vec{\phi}\cdot\bar{\Delta}_\nu\vec{\phi} - \frac{a^2}{6}\left(\bar{\Delta}_\mu\vec{\phi}\cdot\bar{\Delta}_\nu^3\vec{\phi} + \bar{\Delta}_\mu^3\vec{\phi}\cdot\bar{\Delta}_\nu\vec{\phi}\right) + b_2\delta_{\mu\nu}\bar{\Delta}_\rho\vec{\phi}\cdot\bar{\Delta}_\rho\vec{\phi} +$$

$$+ b_3\delta_{\mu\nu}\bar{\Delta}_\mu\vec{\phi}\cdot\bar{\Delta}_\mu\vec{\phi} + b_4\delta_{\mu\nu}m^2\sigma + b_5\delta_{\mu\nu}\vec{\pi}\frac{\delta S_{latt}}{\delta\vec{\pi}}, \tag{67}$$

$$N[\partial_\rho\vec{\phi}\cdot\partial_\rho\vec{\phi}] = c_1\bar{\Delta}_\rho\vec{\phi}\cdot\bar{\Delta}_\rho\vec{\phi} - \frac{a^2}{3}\bar{\Delta}_\rho\vec{\phi}\cdot\bar{\Delta}_\rho^3\vec{\phi} + c_2m^2\sigma + c_3\vec{\pi}\frac{\delta S_{latt}}{\delta\vec{\pi}}. \tag{68}$$

### B. The dilatation current anomaly.

In analogy with the $\lambda\Phi^4$ model, we find that the scale anomaly is proportional to the irrelevant operator

$$a^2\sum_n\left(T_{\mu\mu} - \frac{2}{t}m^2\sigma\right) \tag{69}$$

whose insertions are equivalent to those of

$$a^2\sum_n\left(\beta(t)t\frac{\partial}{\partial t} - \frac{\xi(t)}{2}\vec{\pi}\cdot\frac{\partial}{\partial\vec{\pi}} - \gamma_m(t)m^2\frac{\partial}{\partial m^2}\right)S. \tag{70}$$

Independently of the action (improved or standard) we find

$$\Theta = a^2\sum_n\left\{-\frac{(N-2)}{2\pi}\left[-\frac{1}{2}\bar{\Delta}_\mu\vec{\phi}\cdot\bar{\Delta}_\mu\vec{\phi} + m^2\sigma\right] + \frac{(N-3)}{4\pi}m^2\sigma - t\frac{(N-1)}{4\pi}\vec{\pi}\cdot\frac{\delta S_{latt}}{\delta\vec{\pi}}\right\}, \tag{71}$$

and we obtain the correct one loop (scheme independent) results:

$$\beta(t) = -t\frac{(N-2)}{2\pi}, \tag{72}$$

$$\xi(t) = t\frac{(N-1)}{2\pi}, \tag{73}$$

$$\gamma_m(t) = -t\frac{(N-3)}{4\pi}. \tag{74}$$

### C. Results for the standard theory.

We report the results obtained for the no improved theory. With the same notations used previously we have for the renormalized lagrangian

$$Z_t = 1 + t\frac{(N-2)}{4\pi}\left[\log\frac{32}{4\pi} + \gamma_E\right] + \frac{t}{4}, \tag{75}$$

$$Z = 1 + t\frac{(N-1)}{4\pi}\left[\log\frac{32}{4\pi} + \gamma_E\right]. \tag{76}$$

The relevant insertion of composite operators appearing in $T_{\mu\nu}$ gives

$$b_1 = 1 - \frac{t}{4\pi}\left[4 + \log\frac{32}{4\pi} + \gamma_E\right], \tag{77}$$

$$b_2 = \frac{t}{8\pi}\left[2 - \pi + (N-2)\left(\frac{3}{2} - \log\frac{32}{4\pi} - \gamma_E\right)\right], \tag{78}$$

$$b_3 = -\frac{t}{\pi}\left[1 - \pi\right], \tag{79}$$

$$b_4 = b_5 = t\frac{(N-1)}{8\pi}\left[-\frac{3}{2} + \log\frac{32}{4\pi} + \gamma_E\right], \tag{80}$$

$$c_1 = 1 + \frac{t}{4\pi}\left[\frac{1}{2}(N-2) - (N-1)\left(\log\frac{32}{4\pi} + \gamma_E\right) + 3\pi - 6\right], \tag{81}$$

$$c_2 = c_3 = t\frac{(N-1)}{4\pi}\left[\log\frac{32}{4\pi} + \gamma_E - \frac{1}{2}\right]. \tag{82}$$

The renormalization constants for the energy-momentum tensor are

$$Z_1 = 1 - \frac{t}{4\pi}\left[4 + \pi\right], \tag{83}$$

$$Z_2 = 1 + \frac{t}{4\pi}\left[3\pi - 8 - (N-2)\right], \tag{84}$$

$$Z_3 = \frac{2}{\pi}t\left[1 - \pi\right], \tag{85}$$

$$Z_4 = \frac{t}{8\pi}\left[N - 1\right], \tag{86}$$

$$Z_m = 1 - \frac{t}{8\pi}\left[N - 1\right]. \tag{87}$$

### IV. CONCLUSIONS.

We have defined on the lattice, to one loop, the energy-momentum tensor for two Symanzik tree improved actions, the $\lambda\Phi^4$ scalar theory and the bidimensional nonlinear $\sigma$ model. In the $\sigma$ model case, due to the asymptotic freedom of the theory, the calculation is reliable in the asymptotic scaling regime.

We have found the correct scale anomaly, which as expected is independent of the regularization scheme.

For the $\sigma$ model the corrected EMT we have found can be a starting point for a successive non-perturbative determination with numerical method [18]. As the $T_{00}$ component of EMT is the energy density, it can be used on lattice for the determination of the mass spectrum. We think another interesting possibility is the test of EMT in variational computations, in order to clarify what are in this framework the influences of lattice artefacts.

## APPENDIX A: NOTATIONS AND INTEGRALS.

The lattice propagator is defined as

$$D_{latt}(l) = \frac{1}{\hat{l}^2 + m^2}, \qquad \text{with} \qquad \hat{l}^2 = \sum_{\mu=1}^{D} \frac{4}{a^2} \sin^2\left(\frac{a}{2}l_\mu\right). \tag{A1}$$

We use also

$$\bar{l}_\mu = \frac{1}{a}\sin a l_\mu. \tag{A2}$$

The lattice integral needed for calculations with the standard (not improved) action for $\lambda\Phi^4$ can be evaluated in terms of integrals of modified Bessel functions [15]. All the relevant cases can be expressed in term of two parameters

$$I_1 = \int \frac{d^4l}{(2\pi)^4} \frac{1}{\hat{q}^2 + m^2} = \frac{Z_{0000}}{a^2} - \frac{m^2}{(4\pi)^2}(1 + F_{0000} - \gamma_E - \log a^2m^2) + O(a^2m^4) \tag{A3}$$

$$I_2 = \int \frac{d^4l}{(2\pi)^4} \frac{1}{(\hat{q}^2 + m^2)^2} = \frac{1}{(4\pi)^2}(F_{0000} - \gamma_E - \log a^2m^2) + O(a^2m^2) \tag{A4}$$

$$I_3 = \int \frac{d^4l}{(2\pi)^4} \frac{1}{(\hat{q}^2 + m^2)^3} = \frac{1}{(4\pi)^2}\frac{1}{2m^2} + O(a^2) \tag{A5}$$

$$I_4 = \int \frac{d^4l}{(2\pi)^4} \frac{\bar{q}_\mu \bar{q}_\nu}{(\hat{q}^2 + m^2)^2} = \frac{\delta_{\mu\nu}}{2a^2}\left[\left(Z_{0000} - \frac{1}{8}\right) + \frac{a^2m^2}{(4\pi)^2}(2\pi^2 Z_{0000} + \gamma_E - 1 - F_{0000} + \tag{A6}$$

$$+ \log a^2m^2)\right] + O(a^2m^4) \tag{A7}$$

$$I_5 = \int \frac{d^4l}{(2\pi)^4} \frac{\bar{q}_\mu \bar{q}_\nu}{(\hat{q}^2 + m^2)^3} = \frac{\delta_{\mu\nu}}{4(4\pi)^2}(F_{0000} - 2\pi^2 Z_{0000} - \gamma_E - \log a^2m^2) + O(a^2m^2) \tag{A8}$$

where

$$F_{0000} = \int_0^1 ze^{-8z} I_0^4(2z) dz + \int_1^\infty ze^{-8z}\left[I_0^4(2z) - \frac{e^{8z}}{(4\pi z)^2}\right] \simeq 4.369 \tag{A9}$$

$$Z_{0000} = a^2 \int_{-\pi}^{\pi} \frac{d^4q}{(2\pi)^4} \frac{1}{\hat{q}^2} \simeq 0.155 \tag{A10}$$

For the standard sigma model case all integrals are analitically calculable in term of first and second type elliptic functions.

$$S_1 = \int \frac{d^2 l}{(2\pi)^2} \frac{1}{\hat{l}^2 + m^2} = -\frac{1}{4\pi} \log \frac{a^2 m^2}{32} + O(a^2 m^2) \tag{A11}$$

$$S_2 = \int \frac{d^2 l}{(2\pi)^2} \frac{\cos a l_1}{\hat{l}^2 + m^2} = -\frac{1}{4\pi} \log \frac{a^2 m^2}{32} - \frac{1}{4} + O(a^2 m^2) \tag{A12}$$

$$S_3 = \int \frac{d^2 l}{(2\pi)^2} \frac{\cos a l_1 \cos a l_2}{\hat{l}^2 + m^2} = -\frac{1}{4\pi} \log \frac{a^2 m^2}{32} + \frac{1}{\pi} + O(a^2 m^2) \tag{A13}$$

$$S_4 = \int \frac{d^2 l}{(2\pi)^2} \frac{\cos^2 a l_1}{\hat{l}^2 + m^2} = -\frac{1}{4\pi} \log \frac{a^2 m^2}{32} - \frac{1}{2} + \frac{1}{\pi} + O(a^2 m^2) \tag{A14}$$

The integrals which appear in calculations with the improved actions are not analitically calculable. For the $\lambda \Phi^4$ model we need to extract the asymptotic $a \to 0$ values of

$$M_1 = \int \frac{d^4 l}{(2\pi)^4} D_{latt}(l) = I_1 - \frac{m^2}{(4\pi)^2} T_0 + \frac{Z_0 - Z_{0000}}{a^2} + O(a^2 m^4) \tag{A15}$$

$$M_2 = \int \frac{d^4 l}{(2\pi)^4} D_{latt}^2(l) = I_2 + \frac{1}{(4\pi)^2} T_0 + O(a^2 m^2) \tag{A16}$$

$$M_3 = \int \frac{d^4 l}{(2\pi)^4} D_{latt}^3(l) = I_3 + O(a^2 m^2) \tag{A17}$$

$$M_4 = \int \frac{d^4 l}{(2\pi)^4} D_{latt}^2(l) \overline{q}^2 = \frac{T_1}{a^2} + \frac{2m^2}{(4\pi)^2} (2\pi^2 Z_{0000} + \gamma_E - 1 - T_2 - F_{0000} +$$
$$- \log a^2 m^2) + O(a^2 m^4) \tag{A18}$$

$$M_5 = \int \frac{d^4 l}{(2\pi)^4} D_{latt}^3(l) \overline{q}^2 = \frac{T_2}{(4\pi)^2} + \frac{1}{(4\pi)^2} (F_{0000} - 2\pi^2 Z_{0000} - \gamma_E - \log a^2 m^2) + O(a^2 m^2) \tag{A19}$$

$$M_6 = a^2 \int \frac{d^4 l}{(2\pi)^4} D_{latt}^2(l) \overline{q}^4 = \frac{T_3}{a^2} - \frac{2m^2}{(4\pi)^2} T_4 + O(a^2 m^4) \tag{A20}$$

$$M_7 = a^2 \int \frac{d^4 l}{(2\pi)^4} D_{latt}^3(l) \overline{q}^4 = \frac{1}{(4\pi)^2} T_4 + O(a^2 m^2). \tag{A21}$$

The general strategy is to add and subtract integral resulting from the standard lattice formulation from the improved one. In this way we obtain a divergent piece that can be calculated explicitly (as is a standard $I_i$ integral), and a finite expression that can be evaluated numerically. We give the values of the relevant constants in Tab. I. For the sigma model we need

$$\tilde{S}_1 = \int \frac{d^2 l}{(2\pi)^2} D_{latt}(l) = \frac{1}{4\pi} \left( E_0 - \log \frac{a^2 m^2}{32} \right) + O(a^2) \tag{A22}$$

$$\tilde{S}_2 = \int \frac{d^2 l}{(2\pi)^2} D_{latt}(l) \cos a l_1 = \frac{1}{4\pi} \left( E_5 - \pi - \log \frac{a^2 m^2}{32} \right) + O(a^2) \tag{A23}$$

$$\tilde{S}_3 = \int \frac{d^2 l}{(2\pi)^2} D_{latt}(l) \cos a l_1 \cos a l_2 = \frac{1}{4\pi} \left( E_1 + 4 - \log \frac{a^2 m^2}{32} \right) + O(a^2) \tag{A24}$$

$$\tilde{S}_4 = \int \frac{d^2 l}{(2\pi)^2} D_{latt}(l) \cos^2 a l_1 = \frac{1}{4\pi} \left( E_3 + 4 - 2\pi - \log \frac{a^2 m^2}{32} \right) + O(a^2) \tag{A25}$$

$$\tilde{S}_5 = \int \frac{d^2 l}{(2\pi)^2} D_{latt}(l) \sin^4 a l_1 = \frac{1}{4\pi} E_4 + O(a^2) \tag{A26}$$

$$\tilde{S}_6 = \int \frac{d^2l}{(2\pi)^2} D_{latt}(l) \sin^4 al_1 \cos al_1 \cos al_2 = \frac{1}{4\pi} E_2 + O(a^2) \qquad (A27)$$

$$\tilde{S}_7 = \int \frac{d^2l}{(2\pi)^2} D_{latt}^2(l) = \frac{1}{4\pi m^2} + O(a^2) \qquad (A28)$$

which have been evaluated with the same technique, in terms of the constants reported in Tab. II

TABLES

| | | | |
|---|---|---|---|
| $Z_0$ | 0.129 | $\frac{1}{(4\pi)^2}T_0$ | -0.008 |
| $T_1$ | 0.045 | $\frac{1}{(4\pi)^2}T_2$ | -0.005 |
| $T_3$ | 0.029 | $\frac{T_4}{(4\pi)^2}$ | 0.009 |

TABLE I. The constants $Z_0$ and $T_i$

| | | | |
|---|---|---|---|
| $E_0$ | -0.5928 | $E_1$ | 0.0245 |
| $\frac{1}{4\pi}E_2$ | 0.0443 | $E_3$ | -0.3127 |
| $\frac{1}{4\pi}E_4$ | 0.1056 | $E_5$ | -0.0577 |
| $E_6$ | 0.8706 | $E_7$ | 7.041 |

TABLE II. The constants $E_i$

FIGURES

FIG. 1. The relevant insertions of $T_{\mu\nu}$.

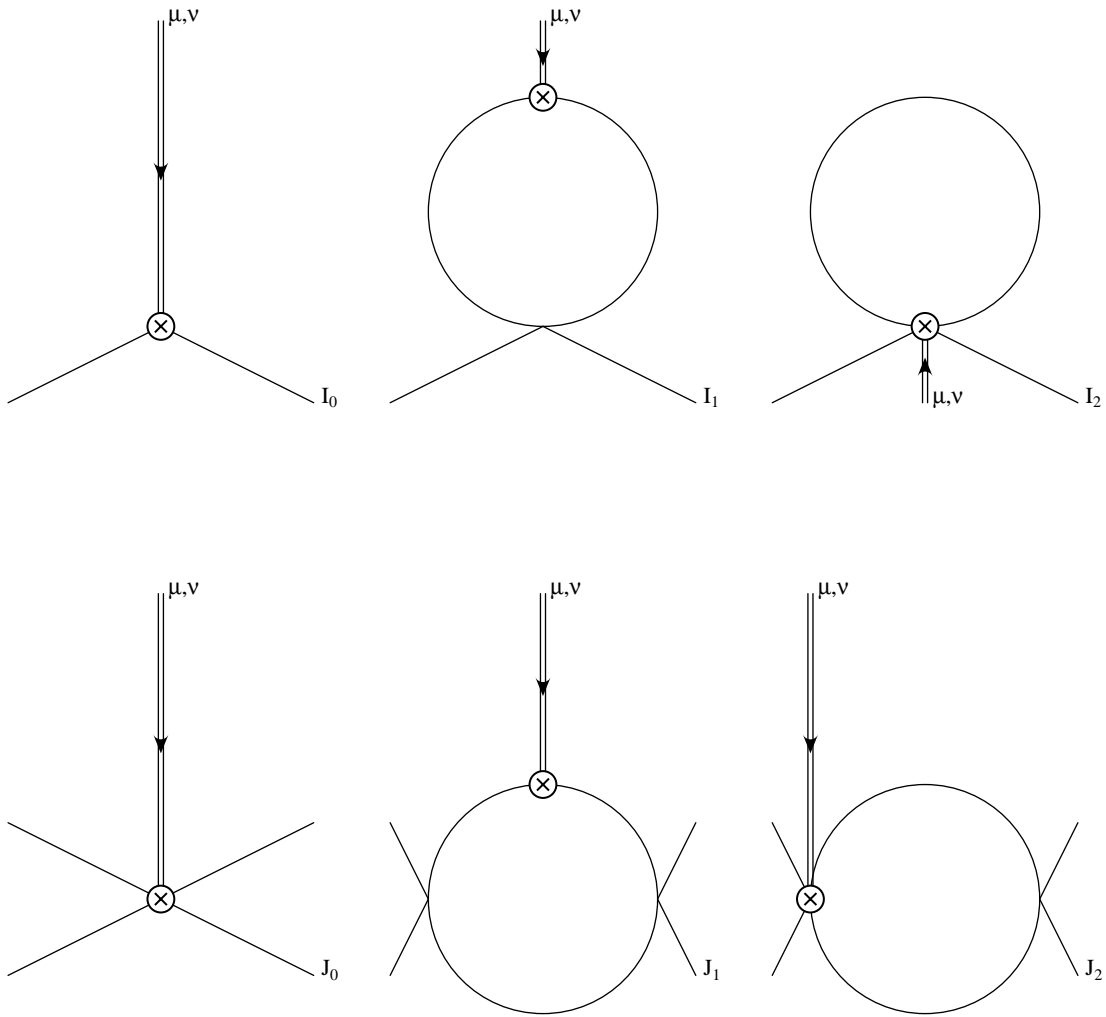

17